\documentclass[smallextended]{svjour3}%
\pdfoutput=1
\usepackage{graphics,amsmath,amsfonts,amscd,revsymb,latexsym,
enumerate,multirow,epsfig}
\usepackage{amsmath}
\usepackage{amsfonts}
\usepackage{amssymb}
\usepackage{graphicx}%
\usepackage{fullpage}
\providecommand{\U}[1]{\protect\rule{.1in}{.1in}}
\providecommand{\U}[1]{\protect\rule{.1in}{.1in}}
\def\bi{\begin{itemize}}
\def\ei{\end{itemize}}
\def\be{\begin{equation}}
\def\ee{\end{equation}}
\def\bea{\begin{eqnarray}}
\def\eea{\end{eqnarray}}
\def\ben{\begin{eqnarray*}}
\def\een{\end{eqnarray*}}

\def\>{\rangle}
\def\<{\langle}

\newcommand{\1} I

\def\*{\star}

\def\0{{\mathbf{0}}}
\def\1{{\mathbf{1}}}
\def\2{{\mathbf{2}}}
\def\3{{\mathbf{3}}}
\def\4{{\mathbf{4}}}
\def\5{{\mathbf{5}}}
\def\6{{\mathbf{6}}}
\def\7{{\mathbf{7}}}
\def\8{{\mathbf{8}}}
\def\9{{\mathbf{9}}}

\smartqed
\journalname{Quantum Information Processing}

\begin{document}
\title{Nonlocal quantum information in bipartite quantum error correction\thanks{M.M.W.
acknowledges support from the MDEIE (Qu\'{e}bec) PSR-SIIRI international
collaboration grant.} }
\author{Mark M. Wilde \and David Fattal}
\institute{Mark M. Wilde is a postdoctoral fellow with the School of Computer Science, McGill University, Montreal, Quebec, Canada H3A 2A7 \email{mark.wilde@mcgill.ca}
\and David Fattal is with the
Information and Quantum Systems Laboratory, Hewlett-Packard Laboratories, 1501
Page Mill Road, MS 1123, Palo Alto, California, 94304-1100, USA
}
\date{Received: \today / Accepted: }


\maketitle
\begin{abstract}
We show how to convert an arbitrary stabilizer code into a bipartite quantum
code. A bipartite quantum code is one that involves two senders and one
receiver. The two senders exploit both nonlocal and local quantum resources to
encode quantum information with local encoding circuits. They transmit their
encoded quantum data to a single receiver who then decodes the transmitted
quantum information. The nonlocal resources in a bipartite code are ebits and
nonlocal information qubits and the local resources are ancillas and local
information qubits. The technique of bipartite quantum error correction is
useful in both the quantum communication scenario described above and in
fault-tolerant quantum computation. It has application in fault-tolerant
quantum computation because we can prepare nonlocal resources offline and
exploit local encoding circuits. In particular, we derive an encoding circuit
for a bipartite version of the Steane code that is local and additionally
requires only nearest-neighbor interactions. We have simulated this encoding
in the CNOT\ extended rectangle with a publicly available fault-tolerant simulation
software. The result is that there is an improvement in the \textquotedblleft
pseudothreshold\textquotedblright\ with respect to the baseline Steane code,
under the assumption that quantum memory errors occur less frequently than
quantum gate errors.


\end{abstract}

\section{Introduction}

Quantum error correction is the theory upon which future quantum computation
and quantum communication devices will depend for reliable operation
\cite{PS95,DG97thesis,CRSS98,qecbook}. This theory is the formal
\textquotedblleft quantization\textquotedblright\ of the classical error
correction theory \cite{FJM77} and borrows several fundamental concepts from
the classical theory, such as digitization, redundancy, and measurement of
errors. Though, the way in which these concepts manifest in quantum error
correction is different from the way that they manifest in classical error correction.

The exploitation of several different forms of \textquotedblleft quantum
redundancy\textquotedblright\ \cite{PhysRevA.56.1721}\ has been a crucial
component of progress
\cite{PS95,KLP05,Bacon05a,aliferis:220502,DP05,BDH06,BDH06IEEE,kremsky:012341,arx2008wildeUQCC}%
\ in the theory of quantum error correction. First, Shor realized that we can
obtain the quantum redundancy necessary for a quantum error-correcting code by
entangling information qubits with extra \textquotedblleft
ancilla\textquotedblright\ qubits \cite{PS95}. Here, the ancilla qubits play
the role of the resource for quantum redundancy. Ten years later, Kribs
\textit{et al.} realized that a noisy qubit, a so-called \textit{gauge} qubit,
is useful for quantum redundancy and formulated the theory of operator quantum
error correction (also known as subsystem quantum error correction)
\cite{KLP05}. Other researchers then showed that quantum codes with gauge
qubits improve the noise threshold for reliable fault-tolerant quantum
computation \cite{Bacon05a,aliferis:220502,DP05}. Shortly after the Kribs
\textit{et al}. result, Brun \textit{et al.} realized that an \textit{ebit}, a
bipartite maximally entangled Bell state shared between a sender and receiver,
is useful for quantum redundancy, and they formulated the theory of
entanglement-assisted quantum error correction \cite{BDH06,BDH06IEEE}. Kremsky
\textit{et al.} followed by showing how ancillas and ebits are useful for the
simultaneous transmission of classical and quantum information and formulated
the classically-enhanced theory of quantum error correction
\cite{kremsky:012341,arx2008wildeUQCC}. These latter two theories of quantum
error correction emerged from advances in quantum Shannon theory, in
particular, from the father protocol \cite{DHW03,DHW05RI}\ and the
classically-enhanced quantum communication protocol \cite{DS03}, respectively.

In this paper, we develop a version of the stabilizer formalism for quantum
error correction that we name the
\textit{bipartite stabilizer formalism}.\footnote{Information theorists would typically denote such a scenario as a
\textquotedblleft multiple access coding\textquotedblright\ scenario
\cite{CT91}. This name is typically reserved for more exotic coding structures
such as those used in code division multiple access \cite{book95viterbi}\ or
superposition coding \cite{book02yeung}. We choose the name \textquotedblleft
bipartite quantum error correction\textquotedblright\ to distinguish the
coding structure presented here because it is not as exotic as either of the
former methods (it simply cuts a stabilizer code into two parts).\ }
A\ bipartite quantum error-correcting code is useful in a quantum
communication scenario in which two senders encode quantum information by
exploiting nonlocal resources that they share. They both then transmit this
encoded quantum information to a single receiver who decodes the transmitted information.

We also introduce a new form of quantum redundancy to the theory of quantum
error correction: a \textit{nonlocal information qubit}. The standard form of
a nonlocal information qubit is as follows:%
\[
\left\vert \varphi\right\rangle ^{AB}\equiv\alpha\left\vert 00\right\rangle
^{AB}+\beta\left\vert 11\right\rangle ^{AB},
\]
where the coefficients $\alpha$ and $\beta$ are arbitrary complex coefficients
such that $\left\vert \alpha\right\vert ^{2}+\left\vert \beta\right\vert
^{2}=1$ and the superscripts indicate that one party, Alice, possesses the
first qubit and another party, Bob, possesses the second qubit. We discuss
this resource in more detail later, but suffice it for now to say that its
power stands somewhere in between an information qubit (or logical qubit) and
an ebit. In this sense, it is perhaps most similar to a coherent bit channel
\cite{Har03,wilde:060303,wilde:022321}. Nonlocal information qubits may arise
naturally in the setting of quantum network protocols \cite{LOW06,KGNR09}\ or
distributed quantum computation
\cite{1997quant.ph..4012G,PhysRevA.59.4249,PhysRevA.62.052317,D05,M06}. In the
sections of this paper on quantum communication, we simply assume that two
senders have such a resource available, perhaps distributed to them from some
\textquotedblleft source\textquotedblright\ party, and the senders would like
to exploit it for quantum communication purposes.

We also show how an example of a bipartite quantum code, a variant of the
Steane code, can lead to increased performance in fault-tolerant quantum
computation (under certain assumptions) \cite{Shor96,AGP06,qecbook}. The
increased performance occurs for this code because its encoding circuit is
localized and has fewer error locations than the encoding circuit for the
baseline Steane code. Additionally, the code retains the error-correcting
properties of the stabilizer code from which it is derived, ensuring the
ability to correct errors on all qubits (similar to the example in
Ref.~\cite{prep2007shaw}). In particular, we present a version of the Steane
code \cite{NC00}\ that requires only nearest neighbor interactions among four
qubits for its encoding, assuming that the seven-qubit quantum register is
initialized with three ebits and one information qubit. Simulations then
demonstrate that this version of the Steane code outperforms the baseline
Steane code, under the assumption that quantum memory errors occur less
frequently than quantum gate errors.

The present work represents an extension of the entanglement-assisted
stabilizer formalism. We exploit recent ideas from the structure of
entanglement-assisted quantum codes \cite{arx2008wildeOEA,W09} and
entanglement measures of stabilizer states \cite{FCY04}\ to construct our
bipartite quantum codes. We motivate new ideas for research pursuits in
network quantum Shannon theory \cite{Yard05a}, specifically related to the
multiple access quantum channel
\cite{Yard05a,Winter01,Ahlswede2005137,KW05,YHD05ieee,itit2008hsieh}.

The next section of this paper develops the resource of a nonlocal information
qubit. Section~\ref{sec:multiparty-stabilizer}\ shows how to construct a
two-sender one-receiver quantum code from any stabilizer code, under the
assumption that the two senders possess ebits and nonlocal information qubits.
That section also includes an example of this type of code. We then present
our fault-tolerant quantum computation simulation results for the performance
of the CNOT extended rectangle \cite{AGP06}\ using a bipartite variation of
the Steane code \cite{NC00}. We finally conclude with some open questions for investigation.

\section{Nonlocal Quantum Information}

\subsection{Nonlocal Information Qubits}

\label{sec:NLI}A nonlocal information qubit is the following state of two
spatially separated qubits:%
\begin{equation}
\left\vert \varphi\right\rangle ^{AB}\equiv\alpha\left\vert 00\right\rangle
^{AB}+\beta\left\vert 11\right\rangle ^{AB}, \label{eq:nonlocal-info-qubit}%
\end{equation}
where one party, Alice, possesses the first qubit and another party, Bob,
possesses the second qubit. We take the standard form of the nonlocal
information qubit to be as above. It is always possible to manipulate the
nonlocal information qubit to be in the above standard form by means of local
operations. Neither Alice nor Bob alone can do anything much with the nonlocal
information qubit, but a third party Claire can decode it if both Alice and
Bob each send her their respective qubit. Claire decodes simply by applying a
CNOT\ from the first qubit to the second qubit once she receives both of them.

The nonlocal information qubit admits mathematical manipulation through the
use of the stabilizer formalism \cite{DG97thesis}. The Pauli operator
$Z^{A}Z^{B}$ stabilizes the nonlocal qubit because the state in
(\ref{eq:nonlocal-info-qubit}) is in the +1-eigenspace of the operator
$Z^{A}Z^{B}$. Local errors of the form $X^{A}$ or $X^{B}$ anticommute with the
operator $Z^{A}Z^{B}$. Claire can detect these errors by measuring the
operator $Z^{A}Z^{B}$ if she possesses both qubits after Alice and Bob
transmit them to her. It is also possible to manipulate the quantum
information in the nonlocal information qubit through the use of logical
operators. The $X$ logical operator of the nonlocal information qubit is
$X^{A}X^{B}$ and the $Z$ logical operator is either $Z^{A}$ or $Z^{B}$. The
logical Hamadard operation is a nonlocal operation that transforms $\left\vert
00\right\rangle ^{AB}$ to $(\left\vert 00\right\rangle ^{AB}+\left\vert
11\right\rangle ^{AB})/\sqrt{2}$ and $\left\vert 11\right\rangle ^{AB}$ to
$(\left\vert 00\right\rangle ^{AB}-\left\vert 11\right\rangle ^{AB})/\sqrt{2}%
$. One cannot fully manipulate the quantum information in the nonlocal
information qubit unless one possesses both qubits or unless, in some cases,
we allow for classical communication between both senders so that, e.g., they
can apply a coordinated $X$ rotation to implement the logical $X$ operator.

We can represent the stabilizer operator $ZZ$\ of a nonlocal information qubit
as the following binary vector by exploiting the Pauli-to-binary isomorphism
(See Ref.~\cite{DG97thesis}):%
\begin{align*}
H  &  \equiv\left[  \left.
\begin{array}
[c]{cc}%
1 & 1
\end{array}
\right\vert
\begin{array}
[c]{cc}%
0 & 0
\end{array}
\right] \\
&  \equiv\left[  \left.
\begin{array}
[c]{cc}%
H_{Z}^{A} & H_{Z}^{B}%
\end{array}
\right\vert
\begin{array}
[c]{cc}%
H_{X}^{A} & H_{X}^{B}%
\end{array}
\right]  ,
\end{align*}
where the matrix to the left of the vertical bar captures the
\textquotedblleft Z\textquotedblright\ part of the operator and the matrix to
the right of the vertical bar captures the \textquotedblleft
X\textquotedblright\ part of the operator. We can represent the logical
operators $XX$ and $ZI$ with the following two respective row vectors:%
\begin{align*}
L  &  \equiv\left[  \left.
\begin{array}
[c]{cc}%
0 & 0\\
1 & 0
\end{array}
\right\vert
\begin{array}
[c]{cc}%
1 & 1\\
0 & 0
\end{array}
\right] \\
&  \equiv\left[  \left.
\begin{array}
[c]{cc}%
L_{Z}^{A} & L_{Z}^{B}%
\end{array}
\right\vert
\begin{array}
[c]{cc}%
L_{X}^{A} & L_{X}^{B}%
\end{array}
\right]  .
\end{align*}
Let $H^{A}$ denote the following matrix:%
\begin{equation}
H^{A}\equiv\left[  \left.
\begin{array}
[c]{c}%
H_{Z}^{A}%
\end{array}
\right\vert
\begin{array}
[c]{c}%
H_{X}^{A}%
\end{array}
\right]  , \label{eq:alice-local-def}%
\end{equation}
and let $H^{B}$, $L^{A}$, and $L^{B}$ denote similarly defined matrices. Let
$G$ denote the following matrix:%
\[
G\equiv%
\begin{bmatrix}
H\\
L
\end{bmatrix}
,
\]
and we can define $G^{A}$ and $G^{B}$ similarly to the definitions of $H^{A}$,
$H^{B}$, $L^{A}$, and $L^{B}$. For any given matrix of the form $F\equiv
\left[  F_{Z}|F_{X}\right]  $, we can define a corresponding symplectic
product matrix $\Omega_{F}$ where%
\begin{equation}
\Omega_{F}\equiv F_{Z}F_{X}^{T}+F_{X}F_{Z}^{T}, \label{eq:symp-prod-mat}%
\end{equation}
and addition is binary. It is straightforward to check that the following
relations hold for the nonlocal information qubit:%
\begin{align*}
\frac{1}{2}\text{rank}\left(  \Omega_{L}\right)   &  =\frac{1}{2}%
\text{rank}\left(  \Omega_{L^{A}}\right)  =\frac{1}{2}\text{rank}\left(
\Omega_{L^{B}}\right) \\
&  =\frac{1}{2}\text{rank}\left(  \Omega_{G}\right) \\
&  =\frac{1}{2}\text{rank}\left(  \Omega_{G^{A}}\right)  =\frac{1}%
{2}\text{rank}\left(  \Omega_{G^{B}}\right) \\
&  =1.
\end{align*}
These types of calculations become important later in this paper because they
allow us to calculate the number of nonlocal information qubits in a given set
of generators.

\subsection{Ebits}

An \textit{ebit} is a special case of a nonlocal information qubit where
$\alpha=\beta=1/\sqrt{2}$. Let $\left\vert \Phi\right\rangle ^{AB}$ denote the
state of an ebit where%
\[
\left\vert \Phi\right\rangle ^{AB}\equiv\frac{\left\vert 00\right\rangle
^{AB}+\left\vert 11\right\rangle ^{AB}}{\sqrt{2}}.
\]
The stabilizer operators of an ebit are $Z^{A}Z^{B}$ and $X^{A}X^{B}$ and,
thus, it has error correction capability only. It can detect local errors of
the form $X^{A}$, $Z^{A}$, $X^{B}$, and $Z^{B}$.

The binary representation of the stabilizer operators of an ebit are as
follows:%
\begin{align*}
H  &  \equiv\left[  \left.
\begin{array}
[c]{cc}%
0 & 0\\
1 & 1
\end{array}
\right\vert
\begin{array}
[c]{cc}%
1 & 1\\
0 & 0
\end{array}
\right]  ,\\
&  \equiv\left[  \left.
\begin{array}
[c]{cc}%
H_{Z}^{A} & H_{Z}^{B}%
\end{array}
\right\vert
\begin{array}
[c]{cc}%
H_{X}^{A} & H_{X}^{B}%
\end{array}
\right]  .
\end{align*}
Defining $H^{A}$, $H^{B}$, $\Omega_{H^{A}}$, and $\Omega_{H^{B}}$ similar to
the way we did in (\ref{eq:alice-local-def}) and (\ref{eq:symp-prod-mat}), it
is straightforward to show that%
\[
\frac{1}{2}\text{rank}\left(  \Omega_{H^{A}}\right)  =\frac{1}{2}%
\text{rank}\left(  \Omega_{H^{B}}\right)  =1.
\]
This result is expected because an ebit contains exactly one ebit of
entanglement and the above matrix rank calculation is equivalent to the
bipartite entanglement measure of Fattal \textit{et al.}~\cite{FCY04}. This
entanglement measure, in turn, coincides with the von Neumann entropy of
entanglement measure.

\subsection{Nonlocal Information Qubits versus Ebits}

The nonlocal information qubit is a hybrid resource for nonlocal quantum
redundancy in a quantum error-correcting code. It mixes the abilities of an
ebit and an information qubit, because it possesses both error detection
capability and information coding ability. That is, it can encode exactly one
qubit into the nonlocal subspace spanned by the states $\{\left\vert
00\right\rangle ^{AB},\left\vert 11\right\rangle ^{AB}\}$, while at the same
time detecting local errors of the form $X^{A}$ and $X^{B}$. In contrast, an
ebit is only useful as an error correction resource because it cannot encode
arbitrary quantum information.

The power of a nonlocal information qubit as used in a quantum
error-correcting code lies in-between that of an information qubit and that of
an ebit (as discussed above). Thus, \textit{there is a qualitative sense} in
which it is similar to a coherent bit channel
\cite{Har03,wilde:060303,wilde:022321}, because a coherent bit channel is a
resource with communication power in-between a noiseless qubit channel and a
shared, noiseless ebit. Consider that the resource of a noiseless information
qubit is qualitatively similar to a noiseless qubit channel because it arises
from the ability to simulate a noiseless qubit channel. That is, there is some
means by which a source $S$\ can distribute an information qubit to Alice with
the noiseless qubit channel%
\[
\left\vert x\right\rangle ^{S}\rightarrow\left\vert x\right\rangle ^{A},
\]
where $x\in\left\{  0,1\right\}  $, if she possesses the noiseless information
qubit $\left\vert \psi\right\rangle ^{A}=\alpha\left\vert 0\right\rangle
^{A}+\beta\left\vert 1\right\rangle ^{A}$. Similarly, the resource of a
nonlocal information qubit is qualitatively similar to a noiseless coherent
bit channel because it arises from the ability to simulate a noiseless
coherent bit channel, where we define a coherent bit channel as the following
isometric map:%
\[
\left\vert x\right\rangle ^{S}\rightarrow\left\vert x\right\rangle
^{A}\left\vert x\right\rangle ^{B},
\]
where $x\in\left\{  0,1\right\}  $. The input system to the coherent bit
channel is a source system $S$, and the output systems are those of Alice and
Bob. In particular, the map maintains coherent superpositions, from which it
gains its name as the coherent bit channel. We leave it open as to who
possesses the source system because it could be Alice (as originally defined
in Ref.~\cite{Har03}) or Bob, or some other system. So, we assume that there
is some means by which a noiseless coherent bit channel is simulated if Alice
and Bob possess a nonlocal information qubit, i.e., there is some means by
which the following map occurs:%
\[
\alpha\left\vert 0\right\rangle ^{S}+\beta\left\vert 1\right\rangle
^{S}\rightarrow\left\vert \varphi\right\rangle ^{AB},
\]
where $\left\vert \varphi\right\rangle ^{AB}$ is defined in
(\ref{eq:nonlocal-info-qubit}). Here, we do not concern ourselves with how
they happen to come upon nonlocal information qubits, but we merely assume
that they have a supply and would like to transmit them to the receiver Claire.

In further analogy of the nonlocal information qubit with the coherent bit
channel, an ebit arises when we send a qubit in the state $\left(  \left\vert
0\right\rangle +\left\vert 1\right\rangle \right)  /\sqrt{2}$ through the
coherent bit channel.

\section{Bipartite Stabilizer Codes}

\label{sec:multiparty-stabilizer}
\begin{figure}
[ptb]
\begin{center}
\includegraphics[
natheight=2.879800in,
natwidth=6.433300in,
height=1.8524in,
width=3.2396in
]%
{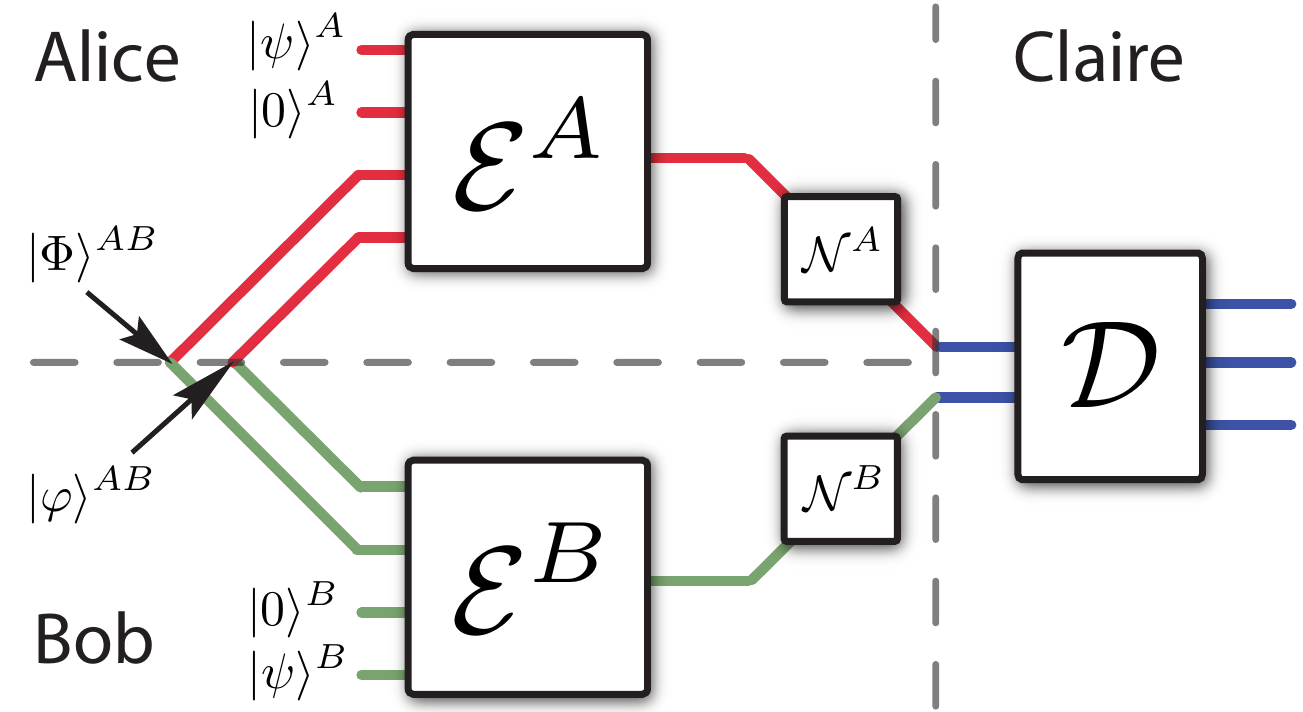}%
\caption{(Color online) The operation of an $\left[  \left[  8,1,1,1;1\right]
\right]  $ two-sender one-receiver quantum error-correcting code. Alice would
like to send the information qubit $\left\vert \psi\right\rangle ^{A}$, Bob
would like to send the information qubit $\left\vert \psi\right\rangle ^{B}$,
and both Alice and Bob would like to send the nonlocal information qubit
$\left\vert \varphi\right\rangle ^{AB}$. For quantum redundancy, Alice uses an
ancilla $\left\vert 0\right\rangle ^{A}$, Bob uses an ancilla $\left\vert
0\right\rangle ^{B}$, and they both exploit an ebit in the state $\left\vert
\Phi\right\rangle ^{AB}$. They both transmit their encoded state over
respective noisy quantum channels $\mathcal{N}^{A}$ and $\mathcal{N}^{B}$
connecting them to Claire, and she then decodes the three information qubits.}%
\label{fig:type1}%
\end{center}
\end{figure}

We now give our model for a bipartite quantum
error correction protocol by constructing a bipartite code from a stabilizer
quantum code. We assume that Alice would like to send $k_{A}$ information
qubits, Bob would like to send $k_{B}$ information qubits, and both Alice and
Bob would like to send $k_{AB}$ nonlocal information qubits. An $\left[
\left[  n,k_{A},k_{B},k_{AB};c_{AB}\right]  \right]  $ two-sender one-receiver
quantum error-correcting code is one that exploits $n$ total channel uses and
$c_{AB}$ ebits shared between Alice and Bob to send the aforementioned amounts
of information qubits. Figure~\ref{fig:type1}\ depicts the operation of
an$\ \left[  \left[  8,1,1,1;1\right]  \right]  $ two-sender one-receiver
quantum code.%

Let us consider a set of commuting generators that forms a valid stabilizer
code \cite{DG97thesis}. Suppose that we have an $\left[  n,k\right]  $
stabilizer code $S$\ with generators $g_{1}$, \ldots, $g_{n-k}$. Suppose that
we also have the $n+k$ generators of the normalizer $N\left(  S\right)  $ of
this code. It is possible to determine the logical operators $\overline{X}%
_{1}$, \ldots, $\overline{X}_{k}$, $\overline{Z}_{1}$, \ldots, $\overline
{Z}_{k}$ of this code by performing a symplectic Gram-Schmidt
orthogonalization procedure on the generators of the normalizer $N\left(
S\right)  $ \cite{W09}---this procedure gives the $2k$ logical operators and
the $n-k$ stabilizer generators.

Now let us assume that Alice possesses the first $n_{A}$ qubits and Bob
possesses the last $n_{B}$ where $n_{A}+n_{B}=n$. We represent the stabilizer
generators as follows:%
\[%
\begin{array}
[c]{ccc}%
g_{1}^{\left(  A\right)  } & \otimes & g_{1}^{\left(  B\right)  }%
\end{array}
,\ldots,%
\begin{array}
[c]{ccc}%
g_{n-k}^{\left(  A\right)  } & \otimes & g_{n-k}^{\left(  B\right)  }%
\end{array}
.
\]
The logical operators are as follows:%
\begin{align*}
&
\begin{array}
[c]{ccc}%
\overline{X}_{1}^{\left(  A\right)  } & \otimes & \overline{X}_{1}^{\left(
B\right)  }%
\end{array}
,\ldots,%
\begin{array}
[c]{ccc}%
\overline{X}_{k}^{\left(  A\right)  } & \otimes & \overline{X}_{k}^{\left(
B\right)  }%
\end{array}
,\\
&
\begin{array}
[c]{ccc}%
\overline{Z}_{1}^{\left(  A\right)  } & \otimes & \overline{Z}_{1}^{\left(
B\right)  }%
\end{array}
,\ldots,%
\begin{array}
[c]{ccc}%
\overline{Z}_{k}^{\left(  A\right)  } & \otimes & \overline{Z}_{k}^{\left(
B\right)  }%
\end{array}
,
\end{align*}
where the $A$ superscript indicates the part of the generator corresponding to
qubits that Alice possesses, and the $B$ superscript indicates the part
corresponding to qubits that Bob possesses. We can alternatively represent the
stabilizer generators with an $\left(  n-k\right)  \times2n$ binary check
matrix $H$ where%
\begin{equation}
H\equiv\left[  \left.
\begin{array}
[c]{c}%
H_{Z}%
\end{array}
\right\vert
\begin{array}
[c]{c}%
H_{X}%
\end{array}
\right]  , \label{eq:first-bin-mat}%
\end{equation}
and the vertical bar divides the matrix into a \textquotedblleft
Z\textquotedblright\ part and an \textquotedblleft X\textquotedblright\ part
according to the Pauli-to-binary isomorphism (See Ref.~\cite{DG97thesis}). We
can further divide the matrix $H$ into matrices corresponding to generators
acting on Alice and Bob's respective qubits:%
\begin{equation}
H\equiv\left[  \left.
\begin{array}
[c]{cc}%
H_{Z}^{A} & H_{Z}^{B}%
\end{array}
\right\vert
\begin{array}
[c]{cc}%
H_{X}^{A} & H_{X}^{B}%
\end{array}
\right]  ,
\end{equation}
where $H_{Z}^{A}$ and $H_{X}^{A}$ are each $\left(  n-k\right)  \times n_{A}$
binary matrices and $H_{Z}^{B}$ and $H_{X}^{B}$ are each $\left(  n-k\right)
\times n_{B}$ binary matrices. Let $H^{A}$ denote the following matrix:%
\begin{equation}
H^{A}\equiv\left[  \left.
\begin{array}
[c]{c}%
H_{Z}^{A}%
\end{array}
\right\vert
\begin{array}
[c]{c}%
H_{X}^{A}%
\end{array}
\right]  ,
\end{equation}
and let $H^{B}$ denote the following matrix:%
\begin{equation}
H^{B}\equiv\left[  \left.
\begin{array}
[c]{c}%
H_{Z}^{B}%
\end{array}
\right\vert
\begin{array}
[c]{c}%
H_{X}^{B}%
\end{array}
\right]  .
\end{equation}
We can represent the normalizer $N\left(  S\right)  $ with an $\left(
n+k\right)  \times2n$ binary matrix $G$ where%
\begin{equation}
G\equiv\left[  \left.
\begin{array}
[c]{c}%
G_{Z}%
\end{array}
\right\vert
\begin{array}
[c]{c}%
G_{X}%
\end{array}
\right]  .
\end{equation}
There is a similar subdivided representation of the matrix $G$:%
\begin{equation}
G\equiv\left[  \left.
\begin{array}
[c]{cc}%
G_{Z}^{A} & G_{Z}^{B}%
\end{array}
\right\vert
\begin{array}
[c]{cc}%
G_{X}^{A} & G_{X}^{B}%
\end{array}
\right]  .
\end{equation}
Let $G^{A}$ denote the following matrix:%
\begin{equation}
G^{A}\equiv\left[  \left.
\begin{array}
[c]{c}%
G_{Z}^{A}%
\end{array}
\right\vert
\begin{array}
[c]{c}%
G_{X}^{A}%
\end{array}
\right]  ,
\end{equation}
and let $G^{B}$ denote the following matrix:%
\begin{equation}
G^{B}\equiv\left[  \left.
\begin{array}
[c]{c}%
G_{Z}^{B}%
\end{array}
\right\vert
\begin{array}
[c]{c}%
G_{X}^{B}%
\end{array}
\right]  . \label{eq:last-bin-mat}%
\end{equation}
The rowspace of matrix $H$ is in the rowspace of matrix$\ G$ because the
generators in group $N\left(  S\right)  $ normalize the generators in the
group $S$.

We can formulate several \textit{symplectic product matrices}
\cite{arx2008wildeOEA,W09}\ that are useful for determining the local
anticommutativity in the above generators. Let $\Omega_{H^{A}}$ be the
\textquotedblleft Alice\textquotedblright\ symplectic product matrix
corresponding to Alice's local matrix $H^{A}$:%
\[
\Omega_{H^{A}}\equiv H_{Z}^{A}\left(  H_{X}^{A}\right)  ^{T}+H_{X}^{A}\left(
H_{Z}^{A}\right)  ^{T},
\]
where addition is binary. Let $\Omega_{H^{B}}$, $\Omega_{G}$, $\Omega_{G^{A}}%
$, and $\Omega_{G^{B}}$ denote similar symplectic product matrices
corresponding to matrices $H^{B}$, $G$, $G^{A}$, and $G^{B}$.

We can manipulate the generators of the stabilizer $S$\ into a form more
suitable for representation as a bipartite code (this manipulation is similar
to that of Theorem~1 in Ref.~\cite{FCY04}). We freely abuse terminology by
referring to a subgroup by its generating set. We first separate the
generators in the stabilizer $S$ into two subgroups with generators of the
following forms:%
\begin{align}
S^{\prime}  &  \equiv\left\{
\begin{array}
[c]{ccc}%
g_{i}^{\left(  A\right)  } & \otimes & g_{i}^{\left(  B\right)  }%
\end{array}
\right\}  ,\nonumber\\
S^{B}  &  \equiv\left\{
\begin{array}
[c]{ccc}%
I^{\left(  A\right)  } & \otimes & g_{j}^{\left(  B\right)  }%
\end{array}
\right\}  . \label{eq:first-step}%
\end{align}
It is possible to bring all generators into one of these two subgroups because
the local \textquotedblleft Alice\textquotedblright\ generators of $S^{B}%
$\ are dependent on the local \textquotedblleft Alice\textquotedblright%
\ generators of $S^{\prime}$. We further manipulate the generators to divide
into three subgroups:%
\begin{align}
S^{AB}  &  \equiv\left\{
\begin{array}
[c]{ccc}%
g_{i}^{\left(  A\right)  } & \otimes & g_{i}^{\left(  B\right)  }%
\end{array}
\right\}  ,\nonumber\\
S^{A}  &  \equiv\left\{
\begin{array}
[c]{ccc}%
g_{j}^{\left(  A\right)  } & \otimes & I^{\left(  B\right)  }%
\end{array}
\right\}  ,\nonumber\\
S^{B}  &  \equiv\left\{
\begin{array}
[c]{ccc}%
I^{\left(  A\right)  } & \otimes & g_{p}^{\left(  B\right)  }%
\end{array}
\right\}  , \label{eq:second-step}%
\end{align}
by using the generators in $S^{B}$ to remove any dependence of the local
\textquotedblleft Bob\textquotedblright\ generators of $S^{\prime}$. We
finally bring the generators of $S$ into the following four subgroups by
performing the symplectic Gram-Schmidt orthogonalization procedure (see
Ref.'s~\cite{arx2008wildeOEA,W09}) on the local \textquotedblleft
Alice\textquotedblright\ part of the generators in $S^{AB}$. This last step
further divides the subgroup $S^{AB}$ into two subgroups $S_{\text{E}}^{AB}$
and $S_{\text{NLI}}^{AB}$:%
\begin{align}
S_{\text{E}}^{AB}  &  \equiv\left\{
\begin{array}
[c]{ccc}%
g_{i}^{\left(  A\right)  } & \otimes & g_{i}^{\left(  B\right)  }\\
\overline{g}_{i}^{\left(  A\right)  } & \otimes & \overline{g}_{i}^{\left(
B\right)  }%
\end{array}
\right\}  ,\nonumber\\
S_{\text{NLI}}^{AB}  &  \equiv\left\{
\begin{array}
[c]{ccc}%
g_{j}^{\left(  A\right)  } & \otimes & g_{j}^{\left(  B\right)  }%
\end{array}
\right\}  ,\nonumber\\
S^{A}  &  \equiv\left\{
\begin{array}
[c]{ccc}%
g_{p}^{\left(  A\right)  } & \otimes & I^{\left(  B\right)  }%
\end{array}
\right\}  ,\nonumber\\
S^{B}  &  \equiv\left\{
\begin{array}
[c]{ccc}%
I^{\left(  A\right)  } & \otimes & g_{q}^{\left(  B\right)  }%
\end{array}
\right\}  . \label{eq:final-form}%
\end{align}

The entanglement subgroup $S_{\text{E}}^{AB}$ consists of those generators in
$S^{AB}$ which have a locally anticommuting partner in $S^{AB}$, where the
anticommutativity is with respect to the local \textquotedblleft
Alice\textquotedblright\ part of the generators. We denote a generator in
$S_{\text{E}}^{AB}$ by $g_{i}^{\left(  A\right)  }\otimes g_{i}^{\left(
B\right)  }$ and its locally anticommuting partner by $\overline{g}%
_{i}^{\left(  A\right)  }\otimes\overline{g}_{i}^{\left(  B\right)  }$. The
generators in $S_{\text{E}}^{AB}$ therefore correspond to ebits that Alice and
Bob share before the quantum communication protocol begins.

The nonlocal information subgroup $S_{\text{NLI}}^{AB}$ consists of those
generators in $S^{AB}$ which have no such locally anticommuting partner in
$S^{AB}$. Its locally anticommuting partners are therefore in the normalizer
$N\left(  S\right)  $. The generators in $S_{\text{NLI}}^{AB}$ therefore
correspond to nonlocal information qubits that Alice and Bob share before the
quantum communication protocol begins. The local subgroups $S^{A}$ and $S^{B}$
correspond to ancilla qubits for Alice and Bob.

The following theorem shows how to produce an $\left[  \left[  n,k_{A}%
,k_{B},k_{AB};c_{AB}\right]  \right]  $ two-sender one-receiver quantum code
from any $\left[  \left[  n,k\right]  \right]  $ stabilizer code, and
furthermore, it computes the parameters $k_{A}$, $k_{B}$, $k_{AB}$, and
$c_{AB}$ as a function of the stabilizer group $S$ and the normalizer
$N\left(  S\right)  $ (it actually uses their corresponding binary representations).

\begin{theorem}
\label{thm:multiparty-stab}From any $\left[  \left[  n,k\right]  \right]  $
stabilizer code, we can produce an $\left[  \left[  n,k_{A},k_{B}%
,k_{AB};c_{AB}\right]  \right]  $ two-sender one-receiver quantum code by
choosing Alice to possess $n_{A}$ qubits and choosing Bob to possess $n_{B}$
qubits, where $n_{A}+n_{B}=n$. The two-sender one-receiver quantum code
requires $c_{AB}$ ebits where%
\[
c_{AB}=\frac{1}{2}\text{rank}\left(  \Omega_{H^{A}}\right)  =\frac{1}%
{2}\text{rank}\left(  \Omega_{H^{B}}\right)  .
\]
It transmits $k_{A}$ information qubits for Alice, $k_{B}$ information qubits
for Bob, and $k_{AB}$ nonlocal information qubits where%
\begin{align*}
k_{AB}  &  =\text{rank}\left(  H^{A}\right)  +\text{rank}\left(  H^{B}\right)
+k-n-c_{AB},\\
k_{A}  &  =\frac{1}{2}\text{rank}\left(  \Omega_{G^{A}}\right)  -c_{AB}%
-k_{AB},\\
k_{B}  &  =\frac{1}{2}\text{rank}\left(  \Omega_{G^{B}}\right)  -c_{AB}%
-k_{AB}.
\end{align*}
We can also compute $k_{AB}$ with the following formula:%
\[
k_{AB}=\frac{1}{2}\left(  \text{rank}\left(  \Omega_{G^{A}}\right)
+\text{rank}\left(  \Omega_{G^{B}}\right)  -\text{rank}\left(  \Omega
_{G}\right)  \right)  -2c_{AB}.
\]

\end{theorem}

\begin{proof}
The method of proof is similar to that originally posted in
Ref.~\cite{arx2008wildeOEA}\ and later exploited in Ref.~\cite{W09}. In this
proof, we liberally go back and forth between the binary representation of
Pauli generators and the groups generated by Pauli generators.

The number $c_{AB}$\ of ebits that the code requires is equal to the number of
locally anticommuting pairs in $S$, with respect to either Alice's local part
or Bob's local part. Ref.~\cite{arx2008wildeOEA} shows that we can calculate
the amount of anticommutativity in any set of generators by calculating the
rank of its corresponding symplectic product matrix and dividing by two. Thus,
the number of ebits that the code requires is equal to%
\[
c_{AB}=\frac{1}{2}\text{rank}\left(  \Omega_{H^{A}}\right)  =\frac{1}%
{2}\text{rank}\left(  \Omega_{H^{B}}\right)  .
\]

We now calculate the number $k_{AB}$\ of nonlocal information qubits. The
number of generators in $S$, or equivalently, the number of rows in $H$, is
equal to $n-k$. The size $\left\vert S^{B}\right\vert $\ of the local subgroup
$S^{B}$ is equal to $n-k$ reduced by the size $\left\vert S^{\prime
}\right\vert $ of the subgroup $S^{\prime}$ in (\ref{eq:first-step}). The size
$\left\vert S^{\prime}\right\vert $\ of the subgroup $S^{\prime}$ in
(\ref{eq:first-step}) is equal to rank$\left(  H^{A}\right)  $ because it
consists of all the locally independent generators. Thus, the size $\left\vert
S^{B}\right\vert $ is%
\[
\left\vert S^{B}\right\vert =n-k-\text{rank}\left(  H^{A}\right)  .
\]
A symmetric argument gives that%
\[
\left\vert S^{A}\right\vert =n-k-\text{rank}\left(  H^{B}\right)  .
\]
These results then imply that the size $\left\vert S^{AB}\right\vert $ of the
subgroup $S^{AB}$ in (\ref{eq:second-step}) is%
\begin{align*}
\left\vert S^{AB}\right\vert  &  =n-k-\left\vert S^{A}\right\vert -\left\vert
S^{B}\right\vert \\
&  =\text{rank}\left(  H^{A}\right)  +\text{rank}\left(  H^{B}\right)  +k-n.
\end{align*}
We obtain the number $k_{AB}$\ of nonlocal information qubits, or
equivalently, the size $|S_{\text{NLI}}^{AB}|$ of the nonlocal information
subgroup $S_{\text{NLI}}^{AB}$, by reducing $\left\vert S^{AB}\right\vert $ by
the number $c_{AB}$\ of ebits:%
\begin{align*}
k_{AB}  &  =\left\vert S^{AB}\right\vert -c_{AB}\\
&  =\text{rank}\left(  H^{A}\right)  +\text{rank}\left(  H^{B}\right)
+k-n-c_{AB}.
\end{align*}

We now calculate the number $k_{A}$\ of \textquotedblleft
Alice\textquotedblright\ local information qubits. For this task, we consider
Alice's part $G^{A}$\ of the full normalizer matrix $G$. The anticommutativity
in the generators corresponding to the rows of the matrix $G^{A}$ is all and
only due to ebits, nonlocal information qubits, and Alice's local information
qubits. The anticommutativity from ebits is due to the local \textquotedblleft
Alice\textquotedblright\ generators in the subgroup $S_{\text{E}}^{AB}$, which
are themselves in the rowspace of $G^{A}$. The anticommutativity from nonlocal
information qubits is due in part to the local \textquotedblleft
Alice\textquotedblright\ generators of $S_{\text{NLI}}^{AB}$ and the local
part of the matching locally anticommuting partners in the local normalizer.
Both of these local generators are in the rowspace of $G^{A}$. Alice's local
information qubits have logical operators which contribute to the
anticommutativity as well. We can calculate the overall number of
anticommuting pairs due to ebits, nonlocal information qubits, and
\textquotedblleft Alice\textquotedblright\ local information qubits as
rank$\left(  \Omega_{G^{A}}\right)  /2$:%
\begin{equation}
\frac{1}{2}\text{rank}\left(  \Omega_{G^{A}}\right)  =k_{A}+c_{AB}+k_{AB}.
\label{eq:local-Alice}%
\end{equation}
Reducing the quantity rank$\left(  \Omega_{G^{A}}\right)  /2$ by the number
$c_{AB}$ of ebits and the number $k_{AB}$\ of nonlocal information qubits
produces the formula for $k_{A}$ in the statement of the theorem. A symmetric
argument gives that%
\begin{equation}
\frac{1}{2}\text{rank}\left(  \Omega_{G^{B}}\right)  =k_{B}+c_{AB}+k_{AB},
\label{eq:local-Bob}%
\end{equation}
and, thus, gives the number $k_{B}$ of local information qubits for Bob.

We can also calculate $k_{AB}$ by recalling that rank$\left(  \Omega
_{G}\right)  /2$ captures the number of logical operators of the code:%
\[
\frac{1}{2}\text{rank}\left(  \Omega_{G}\right)  =k_{AB}+k_{A}+k_{B}.
\]
Combining the above equation with the equations (\ref{eq:local-Alice})
and\ (\ref{eq:local-Bob}) and solving for $k_{AB}$ gives the following formula
for $k_{AB}$:%
\[
k_{AB}=\frac{1}{2}\left(  \text{rank}\left(  \Omega_{G^{A}}\right)
+\text{rank}\left(  \Omega_{G^{B}}\right)  -\text{rank}\left(  \Omega
_{G}\right)  \right)  -2c_{AB}.
\]

Finally, it is possible to produce local encoding circuits for the resulting
two-sender one-receiver quantum code by first bringing the stabilizer
generators into the form in (\ref{eq:final-form}) and applying the algorithm
outlined in the appendix of Ref.~\cite{arx2007wilde} to the local parts of the
generators (this algorithm, in turn, derives from the Grassl-R\"{o}tteler
algorithm for encoding quantum convolutional codes \cite{isit2006grassl}).
\end{proof}

\subsubsection{Example of a Bipartite Stabilizer Code}

We now detail an example of an $\left[  \left[  8,1,1,1;3\right]  \right]  $
two-sender one-receiver quantum error-correcting code. Consider the $\left[
\left[  8,3,3\right]  \right]  $ stabilizer code from Grassl's table of
quantum codes \cite{G07}. Its stabilizer generators are as follows:%

\[%
\begin{array}
[c]{cccccccc}%
X & I & Z & I & Y & Z & X & Y\\
I & X & Z & Z & Y & X & Y & I\\
I & I & X & Y & Z & Z & Y & X\\
Z & I & Z & X & I & Y & Y & Z\\
Z & Z & Z & Z & X & Z & Z & X
\end{array}
.
\]

Let us suppose that Alice possesses the first four qubits and Bob possesses
the second four qubits. Inspection of Alice's local generators reveals that
they are an independent set of generators, and the same holds for Bob's local
generators. Thus, the local subgroups $S^{A}$ and $S^{B}$ are empty.

We then perform the symplectic Gram-Schmidt orthogonalization procedure on
Alice's local generators and produce the following set of generators:%
\[%
\begin{array}
[c]{cccccccc}%
X & I & Z & I & Y & Z & X & Y\\
I & I & X & Y & Z & Z & Y & X\\
I & X & Z & Z & Y & X & Y & I\\
Z & I & Y & Z & Z & X & I & Y\\
I & Y & Z & X & Z & I & Z & I
\end{array}
.
\]
Notice that the first two generators form a locally anticommuting pair, the
second two generators form a locally anticommuting pair, and the last
generator does not have an anticommuting partner. Thus, the first four
generators generate the entanglement subgroup, and the last generator
generates the nonlocal subgroup.

Alice and Bob can each then perform local Clifford operations to reduce the
above stabilizer to the following trivial one:%
\[%
\begin{array}
[c]{cccccccc}%
Z & I & I & I & Z & I & I & I\\
X & I & I & I & X & I & I & I\\
I & Z & I & I & I & Z & I & I\\
I & X & I & I & I & X & I & I\\
I & I & Z & I & I & I & Z & I
\end{array}
.
\]
In the above stabilizer, we can plainly see that the first four generators
correspond to two ebits that Alice and Bob share, and the last generator
corresponds to a nonlocal information qubit (recall from Section~\ref{sec:NLI}
that the stabilizer of a nonlocal information qubit is $ZZ$). The operators
acting on the fourth and eighth qubits are the identity for all stabilizer
generators so that Alice can encode one local information qubit and Bob can
encode one local information qubit. The error-correcting properties of the
code are equivalent to the error-correcting properties of the original
stabilizer code.

\section{Applications to Fault-Tolerant Quantum Computation}

The bipartite coding structure outlined in this paper is useful in
fault-tolerant quantum computation \cite{Shor96,AGP06,qecbook}. This
usefulness is due to the following two factors:

\begin{enumerate}
\item Bipartite codes derived from stabilizer codes maintain their
error-correcting properties.

\item The encoding circuit consists of ebit preparations and local encoding
circuits. For the example code in this section, the encoding circuit requires
only nearest-neighbor interactions and has fewer malignant pairs than the
encoding circuit for the baseline Steane code \cite{AGP06}.
\end{enumerate}

In this section, we represent the Steane code \cite{NC00}\ as a bipartite
quantum code and show how this representation gives a simplified, local
encoding circuit. As a result, the simplified encoding circuit affects the
\textquotedblleft pseudothreshold\textquotedblright\ \cite{SCCA06}\ for
fault-tolerant quantum computation with the Steane code, under the assumption
that quantum memory errors do not occur as frequently as quantum gate errors.
We present the results of numerical simulations that demonstrate how the
pseudothreshold improves under certain assumptions.

\subsection{Steane Code as a Bipartite Stabilizer Code}

Let us first recall the stabilizer generators of the Steane code (as presented
in Ref.~\cite{AGP06}):%
\begin{equation}%
\begin{array}
[c]{ccccccc}%
I & I & I & X & X & X & X\\
I & X & X & I & I & X & X\\
X & I & X & I & X & I & X\\
I & I & I & Z & Z & Z & Z\\
I & Z & Z & I & I & Z & Z\\
Z & I & Z & I & Z & I & Z
\end{array}
. \label{eq:steane-stabilizer}%
\end{equation}
We can write this code as a bipartite code by employing
Theorem~\ref{thm:multiparty-stab}. In particular, let us give the first,
second, and fourth qubits to an \textquotedblleft outside\textquotedblright%
\ party, and the third, fifth, sixth, and seventh qubits to an
\textquotedblleft inside\textquotedblright\ party.\footnote{Note that this
distinction between parties is not particularly relevant in fault-tolerant
quantum computation, but we make this distinction in order to appeal to the
coding structure outlined before.} Figure~\ref{fig:steane-circuit}\ makes
this nomenclature of \textquotedblleft inside\textquotedblright\ and
\textquotedblleft outside\textquotedblright\ more clear. This bipartite cut
yields a $\left[  \left[  7,0,1,0;3\right]  \right]  $ bipartite quantum code
by employing the calculations in Theorem~\ref{thm:multiparty-stab}. It encodes
one local information qubit with the help of three ebits shared between the
inside party and the outside party. Note that this code is also a $\left[
\left[  4,1,3;3\right]  \right]  $ entanglement-assisted code. We refer to
this slight variation of the Steane code as the three-ebit
entanglement-assisted Steane code (\textit{3EA Steane code}).

\subsection{Encoding Method}

The advantage of the 3EA\ Steane code is that it is possible to encode it
using only nearest-neighbor interactions on four qubits (if there is a good
source of ebits available). This property is desirable for a fault-tolerant
encoding circuit because the locality property ensures that errors propagate
only to four qubits during encoding, under the assumption that gate errors
occur more frequently than memory errors.%

\begin{figure}
[ptb]
\begin{center}
\includegraphics[
natheight=3.753300in,
natwidth=6.480000in,
height=1.8879in,
width=3.2396in
]%
{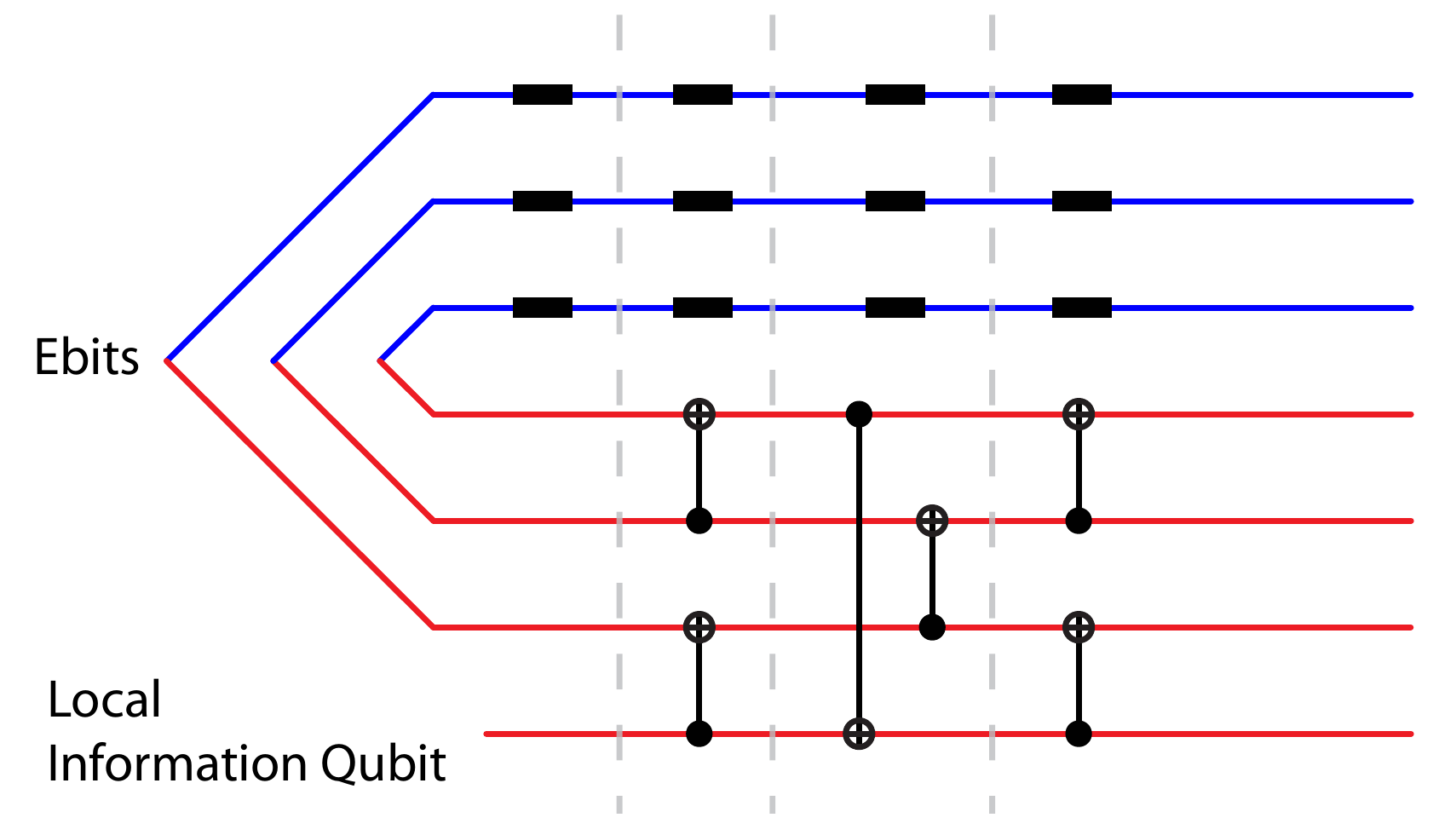}%
\caption{(Color online) The encoding operations for the 3EA\ Steane code. We
first prepare the seven-qubit register with three ebits and one local
information qubit (the red qubits belong to the \textquotedblleft inside
party\textquotedblright\ and the blue qubits belong to the \textquotedblleft
outside party\textquotedblright). The black bars represent \textquotedblleft
wait gates,\textquotedblright\ that model potential memory errors that may
occur. The encoding consists of three rounds of local CNOT\ gates. Note that
we place extra \textquotedblleft memory decoherence\textquotedblright\ on the
ebits to model preparation errors that may occur.}%
\label{fig:steane-encoding-circuit}%
\end{center}
\end{figure}
\begin{figure}
[ptb]
\begin{center}
\includegraphics[
natheight=1.586900in,
natwidth=3.480000in,
height=1.4918in,
width=3.2396in
]%
{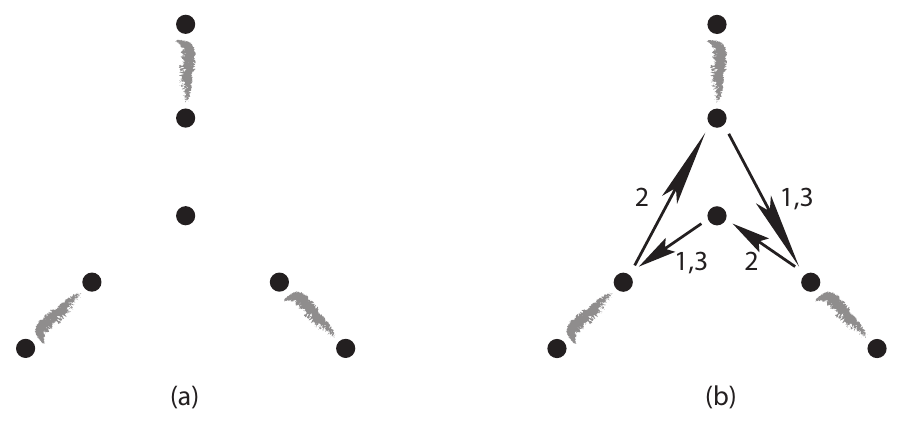}%
\caption{The above figure depicts a particular geometric layout for encoding a
set of qubits with the Steane code. The four inner qubits belong to the
\textquotedblleft inside\textquotedblright\ party and the three outer qubits
belong to the \textquotedblleft outside\textquotedblright\ party. (a) Three
ebits surround an information qubit before the encoding takes place. (b) The
encoding operations are local CNOT\ gates that interact the information qubit
and the three local halves of the ebits. The arrows indicate the direction of
the CNOT gates, and the number adjacent to an arrow indicates the time step at
which the operation takes place.}%
\label{fig:steane-circuit}%
\end{center}
\end{figure}

We now show how to encode the 3EA\ Steane code with local operations. Suppose
that we initialize a seven-qubit quantum register with three ebits and one
information qubit. We assume that the particular technology implementing the
code possesses a good source of ebits, though note that we allow for memory
errors to occur on both halves of the ebits after they have been prepared. The
stabilizer corresponding to the unencoded state is as follows:%
\[%
\begin{array}
[c]{ccccccc}%
I & I & I & X & X & I & I\\
I & X & X & I & I & I & I\\
X & I & I & I & I & I & X\\
I & I & I & Z & Z & I & I\\
I & Z & Z & I & I & I & I\\
Z & I & I & I & I & I & Z
\end{array}
.
\]
The first round of encoding applies a CNOT\ gate from the third qubit to the
fifth and from the sixth qubit to the seventh. The second round of encoding
applies a CNOT\ gate from the seventh qubit to the third and from the fifth
qubit to the sixth. The CNOT gates in the final round are the same as those in
the first round. The result is that the stabilizer of the encoded state is as
given in (\ref{eq:steane-stabilizer}).
Figure~\ref{fig:steane-encoding-circuit}\ depicts the encoding circuit in
quantum circuit notation (with a permutation of the qubits for a simplified
visual presentation). Note that that circuit includes memory errors that may
occur on the outside party's share of the ebits (as explained in the caption
of Figure~\ref{fig:steane-encoding-circuit}). Figure~\ref{fig:steane-circuit}%
\ gives an alternative illustration of the encoding circuit that depicts a
particular geometric layout of the qubits in the seven-qubit quantum register.%

An analysis of this encoding circuit with publicly available fault-tolerant QASM\ tools
\cite{CDT07,C06}\ shows that it has 9,577 CNOT-CNOT\ malignant pairs, which
are quite a bit fewer than the 13,245 CNOT-CNOT\ malignant pairs in the
AGP\ Steane encoding circuit \cite{AGP06}. We then expect that the 3EA\ Steane
code should outperform the baseline Steane code, when quantum memory errors
occur less frequently than quantum gate errors. The results in the next
section confirm this intuition.

\subsection{Simulation Results}%

\begin{figure*}
[ptb]
\begin{center}
\includegraphics[
natheight=3.806000in,
natwidth=4.580000in,
width=6.0534in
]%
{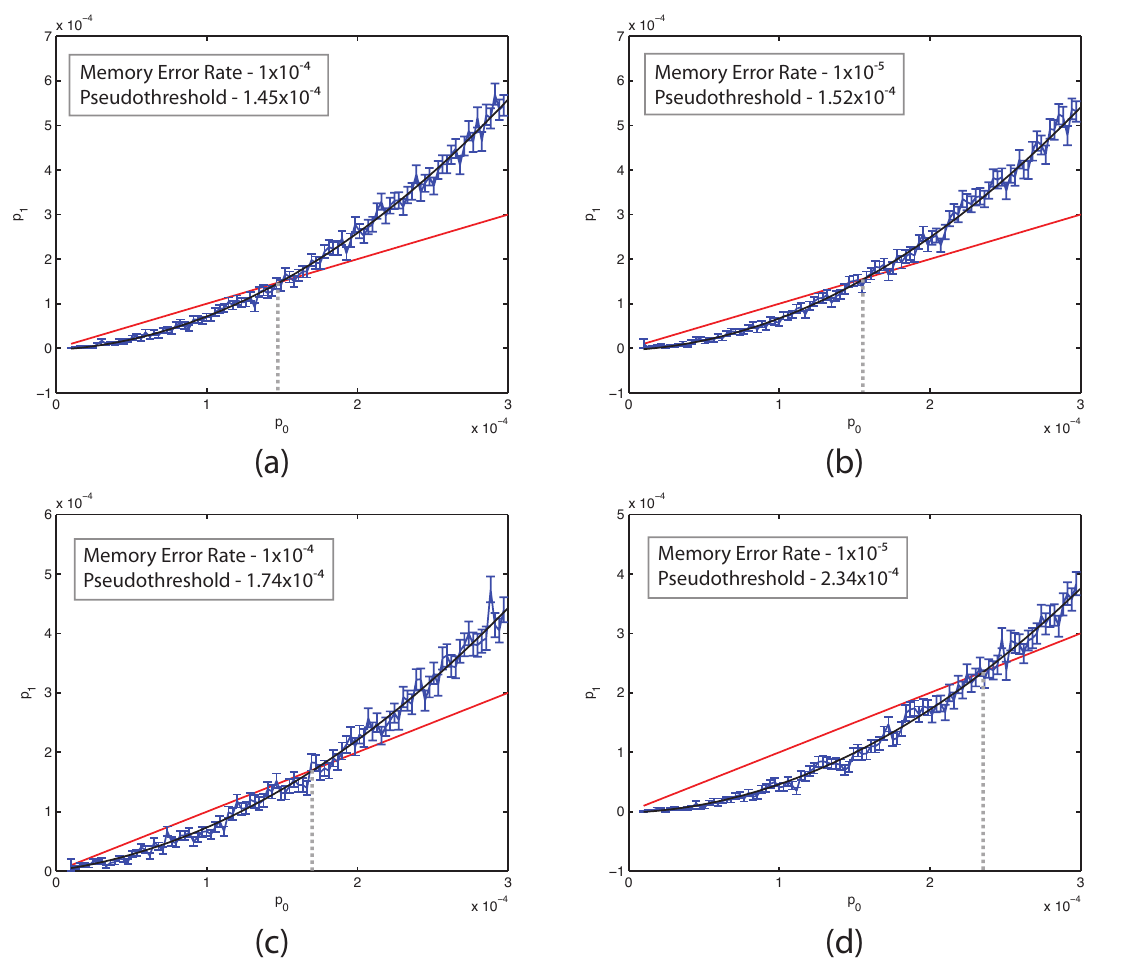}%
\caption{(Color online) Various simulation results for the Steane code in the
CNOT\ extended rectangle. Each plot graphs the probability of failure with
encoding versus the probability of failure without encoding. The red line on
each plot is the probability of failure when the encoding is trivial (i.e.,
there is no encoding). Each blue error bar is the result of $10^{6}$
simulations, and there are 100 blue error bars on each plot. The black curve
going through the error bars is a quadratic fit to the data points. The dashed
line on each plot indicates the location of the pseudothreshold (the point at
which the red curve intersects the black fitted curve). The result is that the
3EA Steane code gives an improvement over the baseline Steane code when memory
errors occur less frequently than gate errors. (a) Baseline Steane code with
memory error rate $1\times10^{-4}$. An estimate of the pseudothreshold is
$1.45\times10^{-4}$. (b) Baseline Steane code with memory error rate
$1\times10^{-5}$. An estimate of the pseudothreshold is $1.52\times10^{-4}$.
(c) 3EA\ Steane code with memory error rate $1\times10^{-4}$. An estimate of
the pseudothreshold is $1.74\times10^{-4}$. (d) 3EA\ Steane code with memory
error rate $1\times10^{-5}$. An estimate of the pseudothreshold is
$2.34\times10^{-4}$.}%
\label{fig:all-data}%
\end{center}
\end{figure*}

We simulated the performance of the 3EA\ Steane code using publicly available
QASM\ fault-tolerant simulation software \cite{CDT07,C06}. In particular, we
evaluated the encoding circuit of the 3EA\ Steane code in the CNOT\ extended
rectangle (See Figure~11\ of Ref.~\cite{AGP06}). The CNOT\ extended rectangle
performs \textquotedblleft Steane error correction\textquotedblright%
\ \cite{PhysRevLett.78.2252} of two logical qubits (recall that
\textquotedblleft Steane error correction\textquotedblright\ is different from
the \textquotedblleft Steane code\textquotedblright), performs a logical
CNOT\ between the two blocks, and performs Steane error correction again on
the two blocks.

Figure~\ref{fig:all-data}\ plots the results of the simulations with
accompanying explanations. The result is that the 3EA\ Steane code gives an
improvement in performance over the baseline Steane code, under the assumption
that quantum memory errors occur less frequently than quantum gate errors.

\section{Conclusion}

The bipartite stabilizer formalism represents a new way of thinking about
quantum error correction codes. Our main theorem shows how to divy up the
qubits in a bipartite quantum code as local information qubits, nonlocal
information qubits, ancilla qubits, and ebits. Our original purpose was to
show how the bipartite stabilizer formalism is useful for quantum
communication, but it turns out to have use in fault-tolerant quantum
computation as well. In particular, the 3EA\ Steane code improves the
pseudothreshold for fault-tolerant quantum computation, under the assumption
that there is a good source of ebits and quantum memory errors occur less
frequently than quantum gate errors. This broad applicability reinforces the
strong links between techniques for quantum communication and techniques for
quantum computation.

We now list several open problems of interest for extensions in quantum
communication topics.

\begin{itemize}
\item It would be interesting to develop quantum Shannon-theoretic protocols
that include nonlocal information qubits. The state merging protocol may be
useful here \cite{nature2005horodecki,cmp2007HOW}.

\item The codes in this paper are relevant for a multiple-access channel, but
it could be interesting to determine if there are useful codes for a broadcast
channel. Yard \textit{et al}. have already explored quantum Shannon theoretic
protocols for the quantum broadcast channel \cite{YHD2006}. Perhaps techniques
from that work will give insight into the design of quantum broadcast channel codes.

\item It might also be interesting to explore bipartite convolutional codes as
an extension of the work in
Ref's.~\cite{arx2007wilde,arx2007wildeEAQCC,arx2008wildeGEAQCC} and the work
in this paper.

\item There might be a way to develop multiparty codes that exploit a common
secret key between multiple parties because of the connection between quantum
privacy and quantum coherence \cite{Devetak03,DW03b,DW03c}.

\item Finally, we are currently considering the extension of the ideas in this
paper to the tripartite setting where the senders share quantum resources with
the receiver (this extension would be similar to the way that
Ref.~\cite{BFG05}\ extends Ref.~\cite{FCY04}).
\end{itemize}

It should also be interesting to explore further improvements that might arise
in fault-tolerant simulations of the 3EA\ Steane when taking into account the
nearest-neighbor interactions in its encoding circuit, similar to the way that
Ref.~\cite{maslov:052310}\ considered the impact of nearest-neighbor
interactions for quantum Fourier transform circuits.

The authors\ thank Todd Brun, Min-Hsiu Hsieh, and Ognyan Oreshkov
for extensive feedback on the manuscript. MMW
acknowledges support from the MDEIE (Qu\'{e}bec) PSR-SIIRI international
collaboration grant.

\end{document}